# THz Generation via Magnetized Hot Plasma-Laser Interaction

M Sharifian and M Arefnia

Atomic and Molecular Group, Department of physics, Yazd University, P.o. Box: 891995-741, Yazd, Iran

Mehdi.sharifian@Yazd.ac.ir

**Abstract**. Terahertz (THz) radiation by beating of two laser beams in a groove density and magnetized hot plasma is discussed theoretically. In this mechanism, the ponderomotive force leads to a nonlinear oscillatory current that resonantly excites THz radiation at the frequency upper hybrid. It is interesting to find that the maximum THz radiation in the various electron temperature, therefore, exited an optimum temperature for the parameters such as beating frequency, transverse distance from lasers beam center and strength of the static magnetic field that THz radiation can be maximized.

## 1. Introduction

One of the promising areas of radiation matter interaction is the study of laser plasma in the frequency ranging between 1 and 10 THz[1-3]. The interaction of pulsed lasers and semiconductors generate THz radiation in photoconductive switching and sampling technique[4-5]. There is a nonlinear interaction of short pulse laser with electro-optic crystals semiconductor/dielectric or synchrotron/transition radiation from bunched electron beams[6-8]. High absorption coefficients of nonlinear optical materials in the THz wave region prevent the optimal efficient THz wave generation[9]. In addition, the non-plasma based techniques encountered as the problem because in high-power laser pulses the material is breakdown. On the other hand, plasma that is already in ionized state has the potential to generate the high-power THz emission[10].

Roskos et al.[11] Amico et al.[12] and Kostin et al.[13] have used high-power laser pulses in many physical mechanisms for THz emissions produced from plasmas. Antonsen et al.[14] examined the ponderomotive force in a plasma from space periodic axial density variation. Recently, Hasanbeigi et al.[15] have investigated the output power and frequency range in presence of plasma can significant increase. Pai et al.[16] beating of two spatial-Gaussian lasers in the presence of a static magnetic field has been used in their novel scheme to produce THz radiation generation. M. Sharifian et al.[17] have discussed the effect of the electron neutral collision frequency on the THz wave generation in the collisional plasma. A. K. Malik et al.[18] have shown that the field of THz radiation is maximized at a specific magnitude of the magnetic field and this maximum energy takes place near the resonance condition. They[19] also have shown that intense, highly focused and tunable THz radiations can be obtained by optimizing the magnetic field, amplitude of ripples, laser profile index, and beam width, etc. Bhasin et al.[20] have studied terahertz radiation in presence of the static electric field and the other researches have considered an azimuthal magnetic field via optical rectification of a laser pulse in a clustered plasma. Pratibha Malik et al.[21] have demonstrated the increment in growth rate, efficiency, and output power occurs by relativistic electron injected into a ripple density plasma. Bakhtiari et al.[22] have investigated mechanism of generating THz radiations by unique features of dark hollow laser beam profile in the collisional plasma.

But, in the works have been done less attention has been paid to the thermal effect, while electron temperature cannot be ignored, especially, for the case of laser-plasma interaction. Zhang et al[23] have discussed the effects of the plasma temperature on the THz radiation in a hot and collisional plasma. They show, high intensity radiation occurs in cold and collisionless plasma and it can be controlled by adjusting the plasma temperature and the Langmuir wave-length

To the best of our knowledge, THz wave characteristics of the hot plasma have been rarely addressed in the plasma literature. In the present study, we have been investigated analytically the effect of the electron temperature on the THz wave generation by beating of two Gaussian co-propagating lasers beam, in the presence of periodically modulated plasma density. The organization of the paper is as follows. First of all, we wrote the equation of motion by considering part of electron temperature, then the nonlinear current that due to the ponderomotive force is obtained. The final normal equation of the terahertz radiation by using phase matching condition and nonlinear current can be achieved. In Section (3), further, we discussed the contribution of electron temperature in various beating frequency, transfer distance from laser beam center and magnetic field for the THz radiation. Conclusions are presented in Section (4).

## 2. Calculation of nonlinear current density

It has been considered two Gaussian lasers of the same intensity, frequencies $\omega_1$ and $\omega_2$ and wavenumber $k_1$ and $k_2$, in the y direction in a grooved plasma having density $n = n_0 + n'$ where $n' = n_\rho e^{i\rho y}$ ($\rho$ is wavenumber and $n_\rho$ is the amplitude of density groove) under the effect of magnetic field B along x-direction. The fields of the lasers are assumed as:

$$\vec{E}_j = \hat{z} E_0 \exp\left[-\frac{z^2}{b_0^2} + i\left(k_j y - \omega_j t\right)\right], \qquad (1)$$

where $j = 1, 2$ and $b_0$ is the initial beam width and z is the transverse distance. Due to the laser fields, the oscillatory velocities electrons achieved by $\vec{v}_j = \frac{e \vec{E}_j}{m i \omega_j}$. Due to spatial variation in laser beam the ponderomotive potential is given by $\varphi_p = -\frac{m}{2e} \vec{v}_1 \vec{v}_2^*$. The ponderomotive force on the electrons at frequency $\omega' = \omega_1 - \omega_2$ and wavenumber $k' = k_1 - k_2$ is equal to:

$$\vec{f}_p^{NL} = -\frac{e^2 E_0^2}{2m\omega_1\omega_2} e^{-2z^2/b_0^2} \left(\hat{y} i k' - \hat{z}\left(\frac{4z}{b_0}\right)\right) e^{i(k'y - \omega' t)}. \qquad (2)$$

By using equation of motion and affected electron temperature, the nonlinear electron velocity $\vec{v}^{NL}$ is obtained as:

$$v^{NL} = -\frac{e\omega^2}{mi\omega(\omega^2 - \omega_c^2)} \nabla \varphi_p + \frac{e\omega_c}{m(\omega^2 - \omega_c^2)} \left(\frac{\partial \varphi_p}{\partial y}\hat{x} - \frac{\partial \varphi_p}{\partial x}\hat{y}\right)$$
$$+ \frac{KT\omega_c}{mn_0(\omega^2 - \omega_c^2)} \left(-\frac{\partial n^{NL}}{\partial y}\hat{x} + \frac{\partial n^{NL}}{\partial x}\hat{y}\right) + \frac{KT\omega^2}{mn_0 i\omega(\omega^2 - \omega_c^2)} \nabla n^{NL}, \qquad (3)$$

(where $\vec{\omega}_c = \frac{eB}{mc}\hat{x}$, $T$ is electron temperature, $K$ is Boltzmann constant), and the equation of continuity is $\frac{\partial n^{NL}}{\partial t} = -n_0 \vec{\nabla} \vec{v}^{NL}$ (where $n_0$ is density of plasma). The ponderomotive force causes the charge separation, therefore, nonlinear perturbation of the density in the presence of the electron temperature is obtained as:

$$n^{NL} = \frac{mn_0}{mi\omega(\omega^2 - \omega_c^2) + k'^2 i\omega KT} \left( i\omega \vec{\nabla} \vec{f}_p^{NL} + \vec{\nabla} \cdot (\vec{f}_p^{NL} \times \vec{\omega}_c) \right), \tag{4}$$

this density perturbation produces an electric field that according to Poisson's equation create a linear density perturbation, given by:

$$n^L = -\frac{\chi_e \vec{\nabla} \cdot (\vec{\nabla}\varphi)}{4\pi e} - \frac{KT \vec{\nabla} \cdot (\vec{\nabla}n^{NL})}{m(\omega^2 - \omega_c^2)}, \tag{5}$$

where $\chi_e = -\frac{\omega_p^2}{(\omega^2 - \omega_c^2)}$ and the plasma frequency is $\omega_p^2 = \frac{4\pi n_0 e^2}{m}$. By using Poisson's equation $\nabla^2 \varphi = 4\pi \tilde{n} e$ with $\tilde{n} = n^{NL} + n^l$ and considering the electron temperature the linear force, due to the self-consistent space-charge field ($\vec{\nabla}\varphi$) is defined as:

$$\vec{f}^L = e\vec{\nabla}\varphi = \frac{\omega_p^2 (\omega^2 - \omega_c^2)}{ie\xi} \left[ i\omega \vec{f}_p^{NL} + (\vec{f}_p^{NL} \times \vec{\omega}_c) \right] - \frac{KT\omega_p^2 (\vec{\nabla}n^{NL})}{en_0(\omega^2 - \omega_h^2)}, \tag{6}$$

where $\xi = \omega(\omega^2 - \omega_c^2)^2(\omega^2 - \omega_h^2) + k'^2\omega(\omega^2 - \omega_c^2)(\omega^2 - \omega_h^2)\frac{KT}{m}$ and $\omega_h^2 = \omega_p^2 + \omega_c^2$ defined as the upper hybrid frequency.

The nonlinear current, $\vec{J}^{NL} = -\frac{1}{2}n'e\vec{v}^{NL}$, due to the presence of nonlinear ponderomotive and linear force, the Lorentz force due to the presence of a magnetic field, and pressure-gradient force is obtained:

$$J^{NL} = -\frac{e^3 n_\rho E_0^2}{4m\omega_1\omega_2} e^{-2z^2/b_0^2} \begin{pmatrix} \frac{\lambda_1 \omega_p^2 \omega_c^2 + \lambda_1^2 \omega^2 + \delta_3 \lambda_2 k'^2 \omega^2}{m\xi} \times \left( k'\hat{y} + i\left(\frac{4z}{b_0^2}\right)\hat{z} \right) \\ + \frac{\lambda_1 \omega_p^2 \omega_c + \lambda_1^2 \omega_c + \delta_3 \lambda_2 k'^2 \omega_c}{m\xi/\omega} \times \left( \left(\frac{4z}{b_0^2}\right)\hat{y} + ik'\hat{z} \right) \\ + \frac{\delta_3 \lambda_1 \omega^2}{m\xi} \times (\delta_1 \hat{y} + i\delta_2 \hat{z}) + \frac{\delta_3 (2\omega_c \omega_p^2 - \lambda_1 \omega_c)}{m\xi/\omega} \times (\delta_2 \hat{y} - i\delta_1 \hat{z}) \end{pmatrix} e^{i(k'y - \omega't)}, \tag{7}$$

where, $\lambda_1 = (\omega^2 - \omega_c^2)$, $\lambda_2 = (\omega^2 - \omega_h^2)$, $\delta_1 = \left( k'^3 + \frac{4}{b_0^2}k' - \frac{16z^2}{b_0^4}k' \right)$, $\delta_2 = \left( -\frac{48z}{b_0^4} - \frac{4z}{b_0^2}k^2 + \frac{64z^3}{b_0^6} \right)$, $\delta_3 = \frac{KT}{m}$.

The current density $\vec{J}^{NL}$ is the source of THz radiation at the frequency $\omega \equiv \omega' = \omega_1 - \omega_2$ and wave number $k \equiv k' + \rho$. The wave equation governing the THz radiation can be written as:

$$-\nabla^2 \vec{E} + \vec{\nabla}(\vec{\nabla} \cdot \vec{E}) = -\frac{4\pi i\omega}{c^2}\vec{J} + \frac{\omega^2}{c^2}\overline{\overline{\varepsilon}}\vec{E}, \tag{8}$$

where $\overline{\overline{\varepsilon}}$ is the plasma permittivity tensor at the THz frequency with $\varepsilon_{xx} = 1 - \frac{\omega_p^2}{\omega^2}$, $\varepsilon_{yy} = \varepsilon_{zz} = 1 - \frac{\omega_p^2}{(\omega^2 - \omega_c^2)}$, $\varepsilon_{zy} = -\varepsilon_{yz} = \frac{i\omega_c \omega_p^2}{\omega(\omega^2 - \omega_c^2)}$, and $\varepsilon_{yx} = \varepsilon_{xy} = \varepsilon_{zx} = 0$. By substituting Eq. (7) and components of permittivity tensor in THz wave equation and with using the concept of the dispersion relation that is $\frac{c^2 k^2}{\omega^2} = 1 - \frac{\omega_p^2 (\omega^2 - \omega_p^2)}{\omega^2 (\omega^2 - \omega_h^2)}$,

we get THz field by $\lambda = \left(\lambda_1 + k'^2 \frac{\lambda_2}{\lambda_1}\right)$ as:

$$\left|\frac{E_{THz}}{E_0}\right| = \left|\frac{in_\rho \delta_3 ey\, \omega \omega_p^2 E_0}{4n_0 mc^2 2k\, \omega_1 \omega_2 (\xi/\lambda_1 \omega)} \left[ \begin{array}{l} (\omega_p^2 \omega_c^2 + \lambda_1 \omega^2)\left(\frac{\omega_c \omega_p^2}{\omega^2 \lambda_2} k' - \frac{4z}{b_0^2}\right) + (\omega_p^2 \omega_c + \lambda \omega_c)\left(\frac{\omega_c \omega_p^2}{\omega \lambda_2}\frac{4z}{b_0^2} - k'\right) \\ + \frac{\omega_c \omega_p^2}{\omega \lambda_2}(\delta_1 - \delta_2) - \left(\frac{2\omega_c \omega_p^2}{\lambda_1} - \omega_c\right)\left(\delta_1 + \frac{\omega_c \omega_p^2}{\omega \lambda_2}\delta_2\right) \end{array} \right] e^{-2z^2/b_0^2} \right|. \quad (9)$$

## 3. Results and discussions

Here the contribution of the electron temperature on the THz radiation in a magnetized hot plasma has been investigated. It would be predicted that the electron temperature has effect on the amplitude of THz radiation.

In order to investigate the importance of the electron temperature, in Figure 1, we analyze the profile of the field amplitude of the emitted THz radiation as a function of the electron temperature whit $\omega_1 = 2.4 \times 10^{14}\, rad/s$, $\omega_p = 2 \times 10^{13}\, rad/s$, $y = 100 c/\omega_p$, $|v_2^*| = 0.3c$, $b_0 = 0.08\, cm$, $z/b_0 = 0.5$, $n_\rho/n_0 = 0.1$. One could see that the field amplitude decreases with an increasing of electron temperature. This is justified physically because of smaller electron density in this process. According to the equations one can see that decreasing of the electron density could decrease the electron current density which plays the main role in the wave equation governing on the THz generation.

In Figure 2, we have plotted THz radiation as a function of normalized beating frequency for two values of electron temperature, 0 and $10^4\, ev$. It can be observed that the THz amplitude achieves maximum value at resonance condition $\omega \approx \omega_h$ (where $\omega_h^2 = \omega_c^2 + \omega_p^2$) then starts decreasing with it frequency. It can also be observed that the field radiation difference of two temperature becomes considerable near the resonance condition.

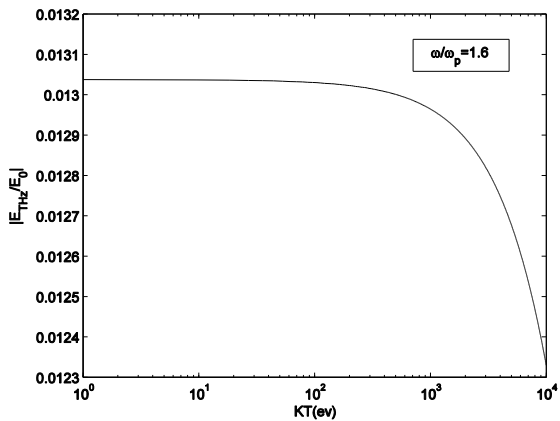

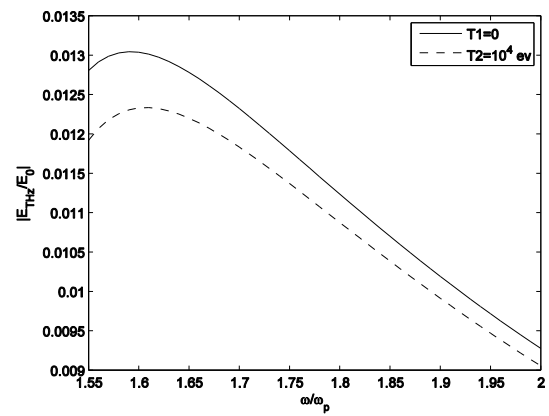

**Figure 1.** THz radiation field vs electron temperature, for $\omega_c = 0.58 \omega_p$

**Figure 2.** Variation in the normalized THz radiation field with normalized beating frequency for two values of electron temperature

We examine the effect of field amplitude radiation as a function of the transverse distance $z/b_0$ for two values of THz electron temperature in Figure 3. We can obtain the radiation of highest intensity desired at near the center of the laser beam. The effect of temperature becomes more significant for the near of the confine than the otherwise.

The effect of the magnetic field for two values of electron temperature is examined for different values of resonance condition $\omega_c/\omega_p$ in Figure 4. It is clear that THz field is maximized at the about $\omega_c/\omega_p \approx 0.6$ (resonance point) when the beating frequency is $\omega/\omega_p = 1.5$. One can evidently see that the effect of the electron temperature on the THz field amplitude is maximum near this resonance point.

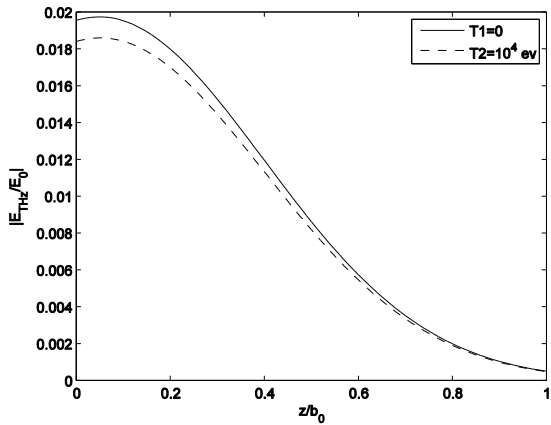 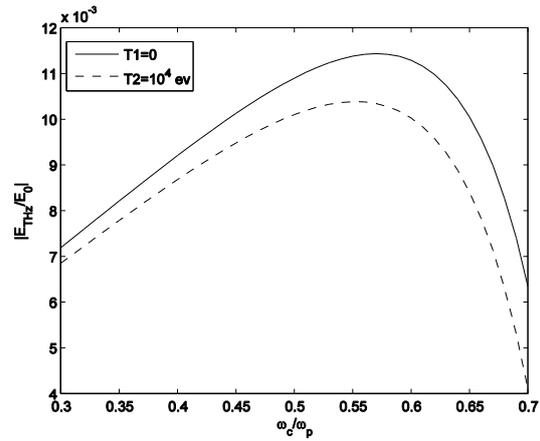

**Figure 3.** The profile of the field amplitude of the emitted THz radiation as a function of the beam width for two values of electron temperature and $\omega/\omega_p = 1.6$

**Figure 4.** Variation in the normalized THz radiation field with normalized electron cyclotron frequency for two values of electron temperature and $\omega/\omega_p = 1.5$

Accordingly, our results explain THz radiation with adjustment parameter of laser and plasma could be controlled by electron temperature.

Figure 5 shows the dependence of THz field on the normalized beating frequency and normalized cyclotron frequency at temperature $10^4 ev$. Evident the peak of the radiation is the near of the resonance condition and the resonance point which is in consistent with presented in Figure 2, 4.

Our surface plot of THz amplitude with variation of beating frequency and electron temperature is shown in Figure 6. It has been found that the radiation peak occurs for lower temperatures, and at resonant condition.

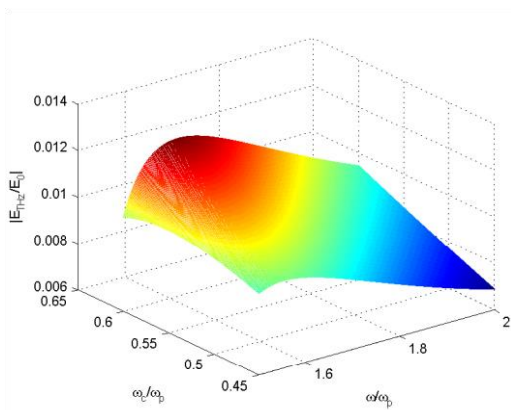 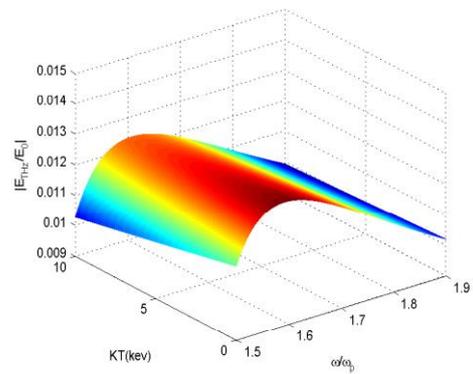

**Figure 5**. Dependence of the normalized THz radiation field on normalized beating frequency and normalized cyclotron frequency when the temperature is $10^4 ev$

**Figure 6**. Dependence of the normalized THz radiation field on normalized beating frequency and electron temperature

## 4. Conclusions

Our analytical calculations show that the generated THz radiation can be affected by the electron temperature. Actually, analytical studies in this work were simultaneous effect of cyclotron frequency and electron temperature ($\frac{KT}{mn_0}(\vec{\nabla} n^{NL})$) in the equation of motion. The field amplitude could be increased with decreasing of electron temperature and this affection depends on the other parameter such as beating frequency, transverse distance from lasers beam center and strength of the static magnetic field. Reduction of the electron temperature is associated with an increase in plasma density, therefore more numbers of electrons take part in the oscillating current is responsible for the emission of THz radiation. It could be found that near the resonance condition, $\omega \approx \omega_h$, the effect of the temperature reduction is strongest. It could be concluded that in the presence of the static magnetic field, temperature dependence of the THz field amplitude is stronger near the center of the laser beam than the otherwise and finally, this effect is maximum at the magnetic resonance point.